\documentclass[12pt]{iopart}
\usepackage{amssymb}
\usepackage{graphicx}
\usepackage{epsfig}
\usepackage{indentfirst}
\usepackage{cite}
\usepackage{subfig}
\usepackage{cleveref}

\begin{document}

\title[Josephson junction based on superconductor/low-resistive normal metal bilayer]{Josephson junction based on highly disordered superconductor/low-resistive normal metal bilayer}

\author{P M Marychev and D Yu Vodolazov}

\address{Institute for Physics of Microstructures, Russian Academy of Sciences, Nizhny Novgorod, 603950 Russia}

\ead{marychevpm@ipmras.ru}

\begin{abstract}
We calculate current-phase relation (CPR) of a SN-S-SN Josephson
junction based on a variable thickness SN bilayer composed of
highly disordered superconductor (S) and low-resistive normal
metal (N) with proximity induced superconductivity. In case when
the thickness of S,N layers and length of S constriction is about
of superconducting coherence length the CPR is single-valued,
could be close to sinusoidal one and the product $I_cR_n$ can
reach $\Delta(0)/2|e|$ ($I_c$ is the critical current of the
junction, $R_n$ is its normal-state resistance, $\Delta(0)$ is the
superconductor gap of single S layer at zero temperature). We
argue that the normal layer should provide good heat removal from
S constriction and there is range of parameters when
current-voltage characteristic is not hysteretic and $I_cR_n$ is
relatively large.
\end{abstract}


\maketitle

\section{Introduction}

Various technological applications of Josephson junctions (AC
voltage standard \cite{Benz-APL-1995}, rapid single-quantum logic
\cite{Likharev-1991}, SQUID-magnetometers \cite{Hazra-PRB-2010}
and particle detectors \cite{Tarte-2000}) require to have
nonhysteretic current-voltage characteristic (IVC). Tunnel
superconductor-insulator-superconductor (S-I-S) junctions is
characterized by small critical current densities and hysteretic
IVC (the latter is related with large capacitance of the insulator
layer) which restricts their applicability. S-N-S and S-S'-S
junctions (where N is a normal metal and S' is the geometric
constriction or superconductor with smaller critical current) have
small capacitance of the weak link but IV curves are hysteretic
due to Joule dissipation
\cite{Hazra-PRB-2010,Skocpol-1974,Courtois_2008,Biswas_2018}.

The current-phase relation (CPR) of these types of the Josephson
junctions is often different from the sinusoidal form
$I=I_c\sin\varphi$ where $I_c$ is the junction critical current
and $\varphi$ is the phase difference between electrodes
\cite{Barone}. Specific form of the CPR depends on the junction
parameters and temperature \cite{CPR-review}. In the case of the
weak link made of pure superconductor or normal metal (having mean
free path $\ell$ much larger than the coherence length of the
electrode $\xi_1$ and the coherence length of the weak link
$\xi_2$), the CPR transform from the sinusoidal one at temperature
close to critical temperature of electrodes $T_c$ to the
saw-toothed shape with the maximum at $\varphi=\pi$ at $T\ll T_c$.
In dirty S-S'-S junctions ($\ell \ll \xi_1$) the CPR with
decreasing temperature can change from the sinusoidal one to the
quite different multi-valued relation. In the latter case the
maximum is attained at $\varphi
>\pi$ and two values of current correspond to a fixed value of
$\varphi$. Similar multi-valued CPR is typical for the weak link
in the form of superconducting bridge whose length is much larger
than the superconducting coherence length $\xi(T)$. In the case of
short bridges (whose length is smaller than the bridge coherence
length $\xi_2$) the CPR remains single-valued at all temperatures
but it is sinusoidal one only at temperature close to critical
temperature and in the case of sufficiently small ratio
$\xi_2/\xi_1$.

For technological applications important characteristic is the
characteristic voltage $V_c=I_cR_n$ where $R_n$ is the
normal-state resistance of the junction. On the one side, to have
large $V_c$ one needs S-N-S or S-S'-S junction with high-resistive
N or S' layer. On the other side, in these junctions IVC becomes
hysteretic below certain temperature which is associated with
Joule heating in the weak link ($\sim I_c V_c \sim I_c^2R_n$) and
the formation at $I>I_c$ of the stable region of suppressed
superconductivity (so called 'hot spot')
\cite{Hazra-PRB-2010,Skocpol-1974,Courtois_2008,Biswas_2018}.
Therefore, the eliminating of the thermal hysteresis without
sacrificing the voltage $V_c$ is important and nontrivial problem.
One solution is a normal metal shunt either on top of the junction
\cite{Lam-2003} or in parallel to it \cite{Muck-1988}. However, in
this case the resistance and the position of the shunt play
important role and they can lead to reduction of the junction
characteristics because of the proximity effect or very small
shunt resistance. In the work \cite{Hadfield-2001} it was proposed
to use as the Josephson junction the variable thickness SN-N-SN
bilayer where the superconducting layer was partially (or
entirely) etched by a focused ion beam. Sufficiently thick
normal-metal layer act as a heat sink which provides nonhysteretic
current-voltage characteristic even at low temperatures. But the
increase of the N layer thickness leads to significant decrease of
$R_n$ and, hence, smaller $V_c$.

In our work we calculate current phase relation for recently
proposed variable thickness SN-S-SN Josephson junction based on
thin dirty superconductor with large normal state resistivity
$\rho_S \gtrsim 100~\mu\Omega\cdot cm$ and thin normal metal layer
with low $\rho_N \gtrsim 2~\mu\Omega\cdot cm$
\cite{Levichev-2019}. In \cite{Vodolazov-2018} it has been
demonstrated theoretically and experimentally that in such a
bilayer the superconducting current mainly flows in N layer (due
to proximity induced superconductivity and $\rho_S/\rho_N \gg 1$)
and the critical current of SN bilayer may exceed the critical
current of single S layer if thicknesses of S and N layers are
about of superconducting coherence length. Below we show that in
comparison with SN-N-SN junction the critical current density
could be about of depairing current density of S layer, which
makes it possible to have $I_cR_n \sim \Delta(0)/2|e|$. Due to
large diffusion coefficient $D_N$ and small minigap in N layer the
heat could be effectively removed from the junction area and
current-voltage characteristic could be not hysteretic. Besides,
because of $D_N\gg D_S$ current-phase relation could be
single-valued at all temperatues and close to sinusoidal one at
temperature near the critical temperature of bilayer.

\section{Model}

\begin{figure}[hbt]
 \begin{center}
\includegraphics[width=0.8\linewidth]{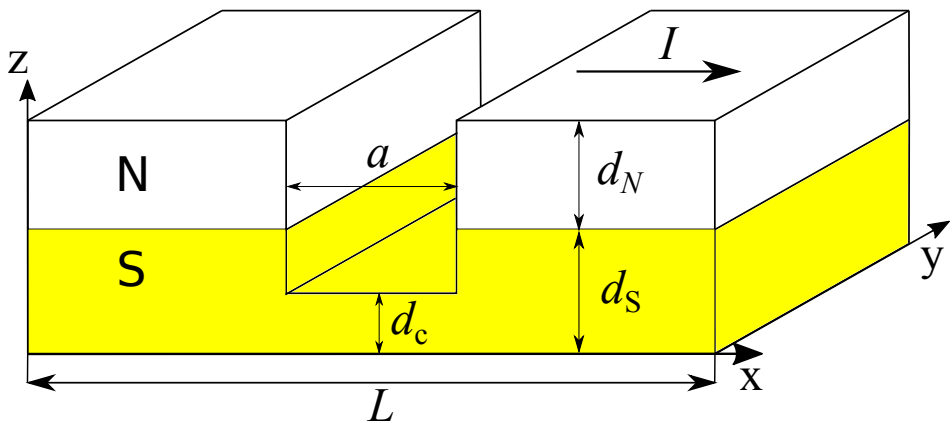}
 \caption{\label{Fig:bridge}
Sketch of SN-S-SN Josephson junction based on variable thickness
SN strip.}
  \end{center}
\end{figure}

The model system consists of SN bilayer strip with length $L$ made
of superconducting film with thickness $d_S$ and normal-metal film
with thickness $d_N$. At the center of bilayer there is a
constriction with length $a$ and thickness $d_c$ where N layer and
partially S layer are removed (see figure \ref{Fig:bridge}). We
assume that in our system the current flows in the $x$ direction
and in the $y$ direction the system is uniform. To find the
current-phase relation of such SN-S-SN Josephson junction at all
temperatures below $T_c$ we solve two-dimensional Usadel equation
for quasiclassical normal $g$ and anomalous $f$ Green functions.
With the angle parametrization $g=cos\Theta$ and
$f=sin\Theta\exp(i\phi)$ this equation in different layers can be
written as

\begin{equation}
 \label{usadel-s}
\frac{\hbar D_S}{2}\left(\frac{\partial^2\Theta_S}{\partial
x^2}+\frac{\partial^2\Theta_S}{\partial
z^2}\right)-\left(\hbar\omega_n+\hbar\frac{D_S}{2}q^2\cos
\Theta_S\right)\sin\Theta_S+\Delta\cos\Theta_S=0,
 \end{equation}
 \begin{equation}
  \label{usadel-n}
  \frac{\hbar D_N}{2}\left(\frac{\partial^2\Theta_N}{\partial
x^2}+\frac{\partial^2\Theta_N}{\partial
z^2}\right)-\left(\hbar\omega_n+\hbar\frac{D_N}{2}q^2\cos
\Theta_N\right)\sin\Theta_N=0,
\end{equation}
where subscripts S and N refer to superconducting and normal
layer, respectively. Here $\hbar \omega_n = \pi k_BT(2n+1)$ are
the Matsubara frequencies ($n$ is an integer number), ${\bf
q}=\nabla\phi =(q_x,q_z)$ is the quantity that is proportional to
supervelocity ${\bf v_s}$, $\phi$ is the phase of superconducting
order parameter. $\Delta$ is the magnitude of order parameter
which should satisfy to the self-consistency equation
\begin{equation}
\label{self-cons-s}
\Delta \ln\left(\frac{T}{T_{c0}}\right)=2\pi k_B
T\sum_{\omega_n
>0} \left(\sin\Theta_S - \frac{\Delta}{\hbar\omega_n}\right),
\end{equation}
where $T_{c0}$ is the critical temperature of the single S layer.
We assume that $\Delta$ is nonzero only in the S layer because of
absence of attractive phonon mediated electron-electron coupling
in the N layer. Equations \eref{usadel-s},\eref{usadel-n} are
supplemented by the Kupriyanov-Lukichev boundary conditions
\cite{Kuprianov-JETP-1988} between layers
\begin{equation}
  \label{boundary-kl}
    \left.D_S\frac{d\Theta_S}{dz}\right|_{z=d_S-0}=\left.D_N\frac{d\Theta_N}{dz}\right|_{z=d_S+0}.
    \end{equation}

In the model we assume transparent interface between N and S
layers which leads to continuity of $\Theta$ on the NS boundary.
At boundaries of the system with the vacuum we use $d\Theta/dn=0$.

To find the phase distribution $\phi$ the equations \eref{usadel-s}~--~\eref{self-cons-s}
are supplemented by two-dimensional equation

\begin{equation}
 \label{divj}
  \textrm{div} {\bf{j_s}}=0,
\end{equation}
where $\bf{j_s}$ is the superconducting current density, which is
determined by the following expression
\begin{equation}
 \label{current}
 {\bf{j_s}}=\frac{2\pi k_BT}{e\rho}{\bf{q}}\sum_{\omega_n > 0}\sin^2\Theta,
\end{equation}
where $\rho$ is the residual resistivity of the corresponding
layer. At the SN-interface we use the boundary condition similar
to (\eref{boundary-kl}) and for the interfaces with the vacuum
we use $d\phi/dn=0$. At the system ends the rigid boundary
conditions are imposed
\begin{equation}
 \label{phase-boundary}
 \phi(0,z)=-\delta\phi/2, \phi(L,z)=\delta\phi/2,
\end{equation}
where $\delta\phi$ is the fixed phase difference between the
system ends. One should differs it with the phase drop near the
junction which we define as
\begin{equation}
 \label{phase}
 \varphi=\delta\phi-kL,
\end{equation}
where $k=q_x (x=0)$ is far from the constriction (in similar way
$\varphi$ is defined in \cite{Baratoff-1970,Zubkov-1983}).
The value of $k$ is found from self-consisting solution of
\eref{usadel-s}~--~\eref{self-cons-s},\eref{divj}.

In numerical calculations we use dimensionless units. The
magnitude of the order parameter is normalized by
$k_BT_{c0}=\Delta(0)/1.76$, length is in units of
$\xi_c=\sqrt{\hbar D_S/k_BT_{c0}}\simeq 1.33~\xi(0)$
($\xi(0)=\sqrt{\hbar D_S/\Delta(0)}$ is the superconducting
coherence length at $T=0$) and current is in units of depairing
current $I_{dep}$ of superconductor at $T=0$.

To calculate the CPR we numerically solve
\eref{usadel-s}~--~\eref{self-cons-s},\eref{divj} by iteration
procedure with fixed $\delta\phi$. When the self-consistency is
achieved (we stop calculations when maximal relative change of $\Delta$
between consequent iterations is less than $10^{-4}$) the Green
functions are used to calculate $j_s$ and the supercurrent per
unit of width $I_s$
\begin{equation}
 \label{sheet-curr}
 I_s=\int\limits_{0}^{d_S+d_N} j_{sx} (x=0) dz.
\end{equation}

We also compare calculated CPR with the current-phase relation for
1D S'-S-S' system with large ratio of diffusion coefficients
$D_{S'}/D_S \gg 1$ (length of S superconductor is equal to $a$).
To calculate it we use 1D Usadel equation.

\section{Current-phase relation of SN-S-SN Josephson junction}

\begin{figure}[hbt]
 \begin{center}
\includegraphics[width=0.96\linewidth]{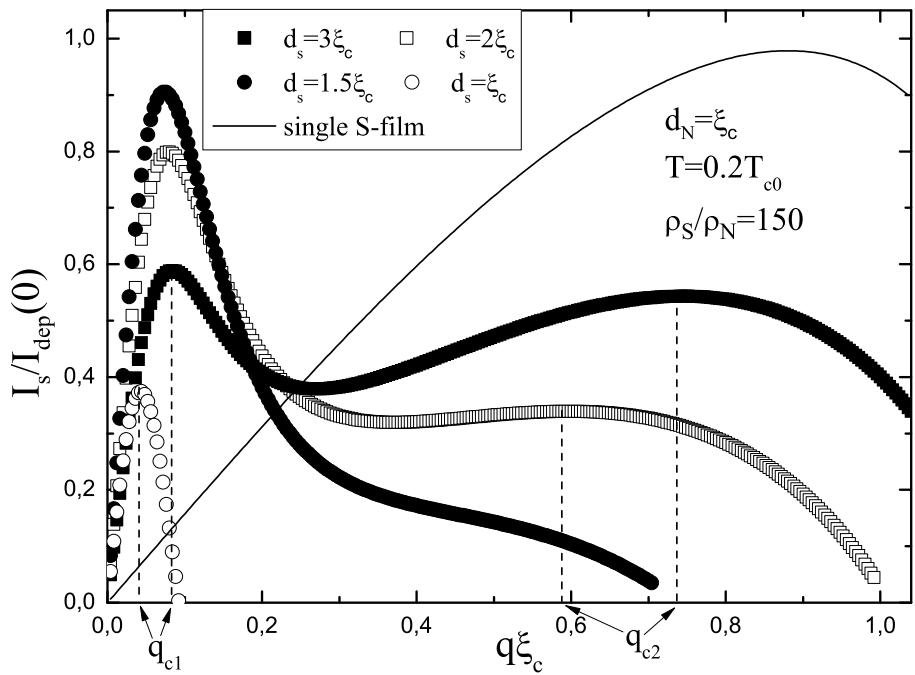}
 \caption{\label{Fig:jq}
Dependence of the superconducting current $I_s$ flowing along SN
bilayer on $q$ for different $d_S$. Solid line shows the
dependence $I_s$ on $q$ for the single S strip. Dashed lines show
the critical values of $q$. Current is normalized by the depairing
current $I_{dep}$ of the single S strip with thickness $d_S$ at
$T=0$.}
 \end{center}
\end{figure}

\begin{figure}[hbt]
 \begin{center}
\includegraphics[width=0.96\linewidth]{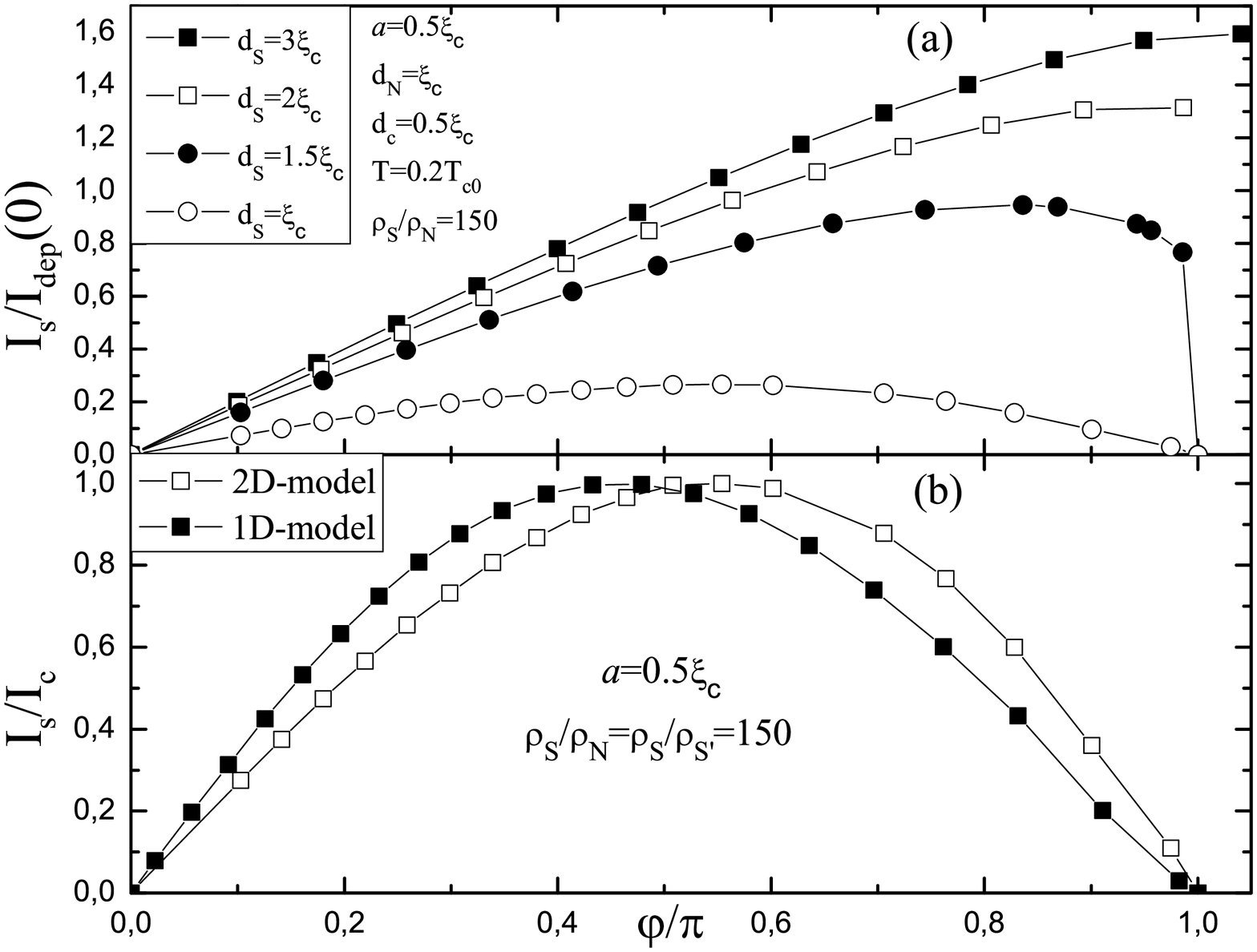}
 \caption{\label{Fig:cpronds}
(a) Current-phase relation of SN-S-SN Josephson junction at
different $d_S$. Current is normalized by the depairing current
$I_{dep}$ of the single S strip with thickness $d_c$ at $T=0$. The
junction parameters are shown in the figure. (b) Comparison of
current-phase relations calculated on the basis of 1D and 2D
models. For 2D case the parameters are following: $d_S=d_N=\xi_c$,
$d_c=0.5\xi_c$, $T=0.2T_{c0}$. In the 1D case temperature
$T=0.6T_{c0}$ corresponds to $T=0.6T_c$, where $T_c = 0.32T_{c0}$
is critical temperature of SN bilayer with chosen parameters. The
superconducting current is normalized by critical current of
Josephson junction.}
 \end{center}
\end{figure}

The dependence $I_s(q)$ in SN bilayer may have one or two maxima
depending on value of $d_S$ (see figure \eref{Fig:jq}) or $d_N$
(see figure 3(a) in \cite{Vodolazov-2018}). The maxima at small
$q$ is connected with suppression of proximity induced
superconductivity in N layer at $q>q_{c1}\sim 1/\sqrt{D_N}$ while
the second maxima at $q=q_{c2}\sim 1/\sqrt{D_S} \gg q_{c1}$ comes
from suppression of superconductivity in S layer when $q>q_{c2}$.
Large difference in $q_{c1}$ and $q_{c2}$ leads to larger phase
concentration in S constriction (see figure \ref{Fig:bridge}) in
comparison with the variable thickness strip (or Dayem bridge)
made of the same material and having the similar geometrical
parameters. Because of that for relatively thin S layers the CPR
is single-valued (see figure \ref{Fig:cpronds} (a)) which is not
easy to achieve for Dayem bridge \cite{Vijay-2009}. For relatively
large $d_S$ there is noticeable contribution to total supercurrent
from S layer which means smaller current (phase) concentration in
constriction like in ordinary Dayem bridge and CPR becomes
multi-valued (see figure \ref{Fig:cpronds}(a) for $d_S=2,3
\xi_c$).

In some respect studied Josephson junction resembles Josephson
junction based on S'-S-S'system composed of two superconductors S
and S' having $D_{S'}\gg D_S$ and the same thicknesses
$d_S=d_{S'}$
\cite{Baratoff-1970,Baratoff-1975,Kupriyanov-LTP-1981}. Josephson
junction based on this quasi 1D system has single-valued CPR which
tends to the sinusoidal shape with increasing temperature. In figure \ref{Fig:cpronds} (b)
we compare CPR calculated for 1D S'-S-S' and 2D SN-S-SN systems.
Since in 1D model there is no suppression of $T_c$ by N layer, in
calculations we use ratio $T/T_{c0}$ which corresponds to ratio
$T/T_c$ of 2D SN structure. Visible differences between CPRs
calculated using different models could be related with
transversal inhomogeneity near the S constriction in the 2D case.
\begin{figure}[hbt]
 \begin{center}
\includegraphics[width=0.96\linewidth]{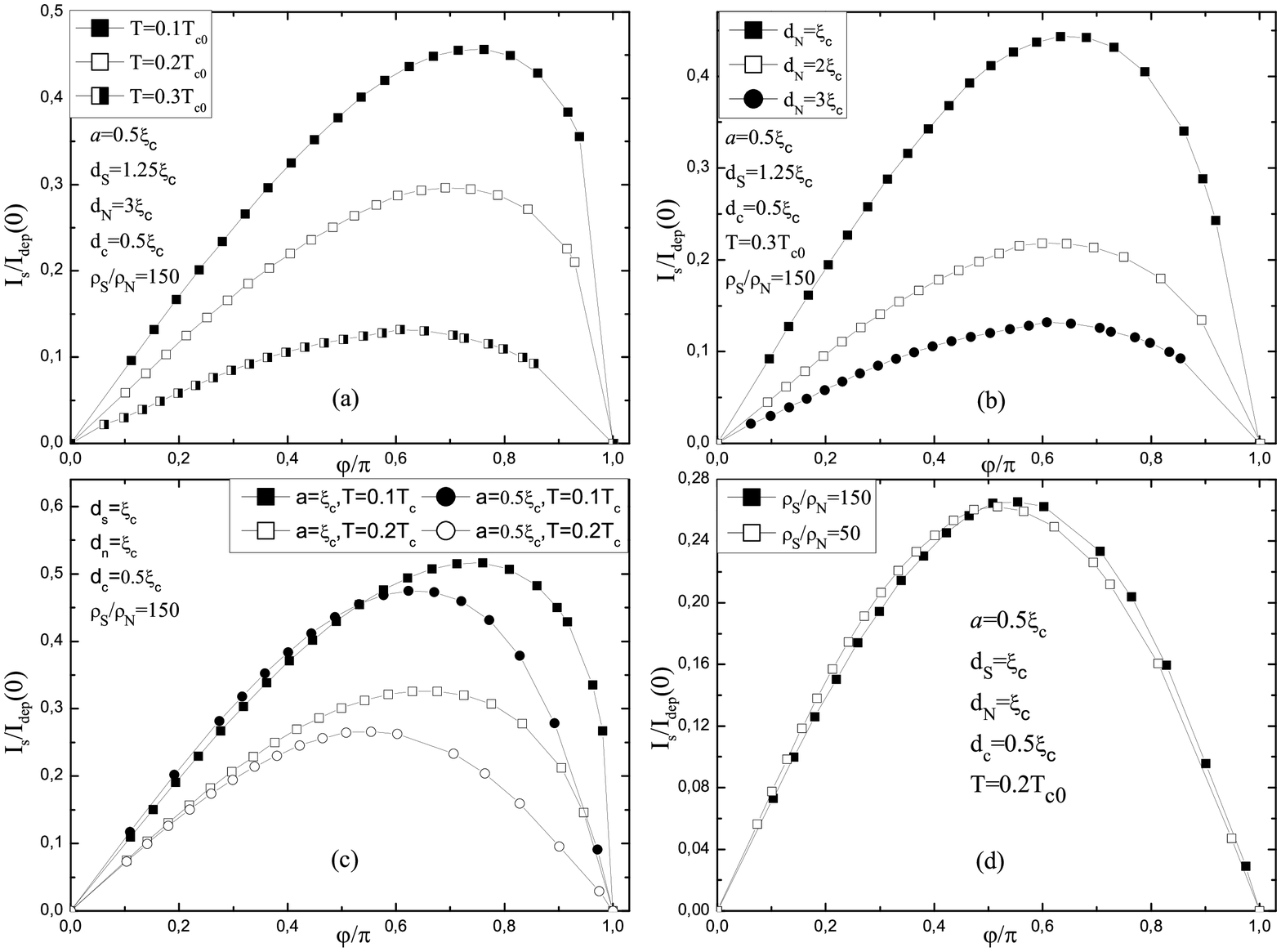}
 \caption{\label{Fig:cprdiffpar}
Variation of current-phase relation of SN-S-SN junction with
change of: (a) temperature; (b) thickness of N layer $d_N$; (c)
length of constriction $a$; (d) ratio of resistivities. Current is
normalized by the depairing current $I_{dep}$ of the
superconducting strip with thickness $d_c$ at $T=0$.}
\end{center}
\end{figure}

We have studied evolution of CPR of SN-S-SN Josephson junction by
varying different parameters. In figure \ref{Fig:cprdiffpar}(a) we
demonstrate that with increase of the temperature the current
phase relation becomes closer to sinusoidal one which is typical
for S'-S-S' junctions \cite{Kupriyanov-LTP-1981} and it is related
with increase of the temperature-dependent coherence length
$\xi(T)$. Effect of different $d_N$ is shown in figure
\ref{Fig:cprdiffpar}(b). An increase in $d_N$ leads to slight
shift of maximum of $I_s(\varphi)$ to the left and decrease of
$I_c$ which are explained by lowering of $T_c$ of SN bilayer for
thicker N layers. Lower $I_c$ means smaller $I_cR_n$ but how we
discuss below large $d_N$ provides better cooling of S
constriction and nonhysteretic IV curves.

An increase of the weak-link length $a$ leads to the shift of the
maximum of $I_s(\varphi)$ to the right (see figure
\ref{Fig:cprdiffpar}(c)) as it is typical for ordinary variable
thickness Josephson junctions. Interestingly, that contrary to
that junctions the $I_c$ increases in SN-S-SN system. This result
is explained by lower value of superconducting order parameter in
SN banks in comparison with $\Delta$ in S constriction at $I_s=0$.
With increasing $a$ the supercondcuting order parameter in
constriction increases and $I_c$ increases too.

And finally figure \ref{Fig:cprdiffpar}(c)) illustrates that
three-fold decrease of ratio $\rho_S/\rho_N$ does not change
current-phase relation drastically. Both the critical current and
shape of CPR vary a little.

\section{Effect of Joule heating in SN-S-SN junctions}

The absence of hysteresis in current-voltage characteristic is
important for devices based on Josephson junctions. The hysteresis
in Dayem, variable thickness, S'-S-S' or S-N-S junctions is mainly
caused by the temperature rise in the weak-link region in the
resistive state due to Joule heating and the formation of hot spot
\cite{Courtois_2008,Hazra-PRB-2010,Biswas_2018}. Local heat
production should be large in SN-S-SN junction due to large
critical current density which is about of the depairing current
density of the superconductor. But as we show below the presence
of relatively thick N layer with large diffusion coefficient
provides efficient cooling of constriction.

To estimate the increase of temperature in the resistive state  we
use two temperature (2T) model \cite{Perrin-1983,Vodolazov_2017}
for SN-S-SN junction. We suppose that electron $T_e=T+\delta T_e$
and phonon $T_p=T+\delta T_p$ temperatures are near the substrate
temperature $\delta T_e, \delta T_e \ll T$ and do not vary along
the thickness. Because of inverse proximity effect the gap in
relatively thin S layer ($d_S \lesssim 1.5 \xi_c$) is suppressed
in comparison with single S layer, which permits heat diffusion
from N to S layer in SN banks. In S constriction being in the
resistive state at $I>I_c$ the superconducting order parameter is
also suppressed. It allows us to use normal state heat
conductivity both in SN and S regions in heat conductance equation
for calculation of $\delta T_e$. In our model Joule dissipation is
taken into account only in S constriction, because in SN bilayer
it is considerably lower due to much lower resistivity and lower
current density. Because of small length of constriction and large
difference in diffusion coefficients and thicknesses in
constriction and banks we can neglect heat flow to the phonons and
substrate in constriction (main cooling of junction comes from
diffusion of hot electrons to SN banks). In SN bilayer $D_N \gg
D_S$ and heat diffusion occurs mainly along N layer. With above
assumptions we have following equation for $\delta T_e$
\begin{eqnarray}
 \label{tempn}
\frac{d^2\delta T_e}{dx^2}+\rho_S(j_c)^2/\kappa_S =0, |x|\leq
a/2,
 \\
 \nonumber
\frac{d^2\delta T_e}{dx^2}- \frac{\delta T_e}{\lambda_T^2} =0,
|x|\geq a/2,
\end{eqnarray}
where $\kappa_S=2D_SN(0)k^2_BT/3$ is the electron heat
conductivity of S layer in the normal state, $N(0)$ is the one
spin density of states on the Fermi level,
\begin{equation}
 \lambda_T=\sqrt{D_N\tau_0}
 \left(\frac{T_{c0}}{T}\right)^{3/2}\sqrt{\frac{\pi^2(1+\beta)}{720 \zeta(5)}}
\end{equation}
is the healing length, $\beta=[\gamma\tau_{esc}
450\zeta(5)T/[\tau_0\pi^4T_{c0}$], $\zeta(5)\simeq 1.03$,
$\tau_{esc}$ is the escape time of nonequilibrium phonons to
substrate, $\gamma=8\pi^2C_e(T_{c0})/C_p(T_{c0})$ is the ratio of
electron and phonon heat capacities at $T=T_{c0}$ and $\tau_0$
determines the strength of electron-phonon inelastic scattering in
S and N layers (see equations (4,6) in \cite{Vodolazov_2017}). For
$\tau_0$ we use the smallest time for S and N materials due to
assumed good transfer of electrons between S and N layers and
their small thickness. On the boundary between S and SN regions we
use continuity of the electron temperature ($\delta
T_e|_{a/2-0}=\delta T_e|_{a/2+0}$) and heat flux
($d_cD_S\frac{\delta T_e}{dx}|_{a/2-0}=d_ND_N \frac{\delta
T_e}{dx}|_{a/2+0}$).

Using (10) and above boundary conditions we find maximal
temperature increase in the constriction

\begin{equation}
\frac{\delta
T_e^{max}}{T}=0.23\left(\frac{a}{\xi_c}\right)^2\left(\frac{T_{c0}}{T}
\right)^2\left(\frac{I_c}{I_{dep}(0)} \right)^2
\left(\frac{D_Sd_c}{D_Nd_N}\frac{4\lambda_T}{a}+1 \right)
\end{equation}

In following estimations we use parameters of NbN (S layer) and Cu
(N layer): $T_{c0}=10$ K, $D_S=0.5$ cm$^2$/s, $\rho_S = 200~\mu
\Omega \cdot$cm, $D_N=40$ cm$^2$/s, $\rho_N= 2~\mu \Omega \cdot$cm,
$\tau_0=1$ ns (theoretical estimation for NbN is taken from \cite{Vodolazov_2017}), $\xi_c=6.4$ nm, $\gamma=9$,
$d_S=1.25\xi_c$, $d_N=2 \xi_c$, $\tau_{esc}=4(d_N+d_S)/u \simeq 41$
ps ($u= 2\cdot 10^5$ cm$^2$/s is a mean speed of sound),
$T/T_{c0}=0.3$, $T_c/T_{c0}=0.43$, $a=0.5\xi_c$, $d_c=0.5\xi_c$. With these
parameters $\beta \simeq 0.53$, $I_c\simeq 0.22 I_{dep}(0)$ (see
figure \ref{Fig:cprdiffpar}(b)) and $\delta T_e^{max}/T \sim 0.24$ is
small, thanks to $D_N \gg D_S$ and $d_N \gg d_c$.

\section{Discussion}

We use Usadel model to calculate current-phase relation of SN-S-SN
Josephson junction based on high-resistive superconductor and
low-resistive normal metal. In \cite{Vodolazov-2018} from
comparison of the experiment and theory it was concluded that
Usadel model underestimates proximity induced superconductivity in
N layer and overestimates inverse proximity effect in S layer in
NbN/Al, NbN/Ag and MoN/Ag bilayers. Namely, the suppression of
critical temperature of SN bilayer is smaller while change in
magnetic field penetration depth of SN bilayer is larger than
Usadel model predicts. Therefore, present results should be
considered only as a route for possible experimental realization
of SN-S-SN Josephson junction. They demonstrate that the thickness
of S layer should not exceed $\sim 1.5 \xi_c$, otherwise
current-phase relation is not single-valued for reasonable length
and thickness of S constriction. The thickness of N layer should
not be too small (small $d_N$ leads to large overheating) and not
too large (the larger $d_N$ leads to lower $T_c$ and smaller $I_c$
at fixed substrate temperature).

Our results show that SN-S-SN Josephson junction in many respects
resembles Dayem, variable thickness bridge, S'-S-S' or S-N-S
junctions. Product
\begin{equation}
V_c=I_cR_n =
\frac{\Delta(0)}{|e|}\frac{a}{\xi_c}\frac{I_c}{I_{dep}(0)},
\end{equation}
can reach $0.5 \Delta(0)/|e|$ at low temperature ($T=0.1 T_{c0}$)
and $a=\xi_c$ (see figure \ref{Fig:cprdiffpar}(c)) due to use of
superconductor in constriction area, instead of normal metal as in
\cite{Hadfield-2001}. In case of NbN with $T_{c0}=10 K$ one may
have $V_c=0.75$ mV but according to (12) $\delta T_e^{max}$ will
be larger than $T$ at these parameters. However there is a hope,
that critical temperature of real SN bilayer is higher than Usadel
model predicts (see discussion above) and therefore large $I_c$
could be reached at higher operating temperature $T/T_{c0}$,
leading to drastic reduction of $\delta T_e^{max}$ (see (12)).

The SN-S-SN junctions made of NbN/Al bilayer have been fabricated
recently \cite{Levichev-2019} and indications of Josephson effect
(the presence of Shapiro steps and Fraunhofer like dependence of
critical current on the magnetic field) have been observed. But
due to not optimized parameters ($d_S=d_c \sim  15$ nm $\sim 2.3
\xi_c$, $d_N \sim 29$ nm $\sim 4.5 \xi_c$, $a=20$ nm $\sim 3.1
\xi_c$) the IV curves were hysteretic already at temperature close
to critical one and width of Shapiro steps did not follow the
theoretical expectations \cite{Levichev-2019}. Modern technology
allows to make constriction with length about $5$ nm with help of
helium beam, which is smaller than $\xi_c$ in NbN. Successful
implementation of this method could lead to creation of low
temperature nano-scale Josephson junction or their arrays. For
example SN-S-SN junctions can be promising to use in programmable
voltage standards \cite{Benz-APL-1995} where large value of $V_c$
allows to reduce the number of junctions and to use Shapiro steps
of order higher than one. Nonhysteretic current-voltage
characteristics with large $V_c$ at low temperatures enables to
use these structures for various low-temperature applications,
e.g., particle detectors \cite{Tarte-2000}.

\section{Conclusion}

In conclusion, we have calculated current phase relation of
Josephson junction based on variable thickness SN-S-SN strip,
where S is dirty superconductor with large normal state
resistivity and N is low resistive normal metal. We find the range
of parameters when CPR is single-valued, close to sinusoidal one
and product $I_cR_n \lesssim \Delta(0)/2|e|$. Our estimations
demonstrate that relatively thick N layer serves as effective
heat-conductor providing weak overheating and nonhysteretic
current-voltage characteristic of SN-S-SN Josephson junction.

\ack

P. M. M. acknowledges support from Russian Scientific Foundation
(project No. 20-42-04415) and D. Yu. V. acknowledges support from
the Foundation for the Advancement of Theoretical Physics and
Mathematics BASIS (program No. 18-1-2-64-2).


\section*{References}

\begin{thebibliography}{20}
\bibitem{Benz-APL-1995} Benz S P 1995 \textit{Appl. Phys. Lett.} {\bf 67} 2714

\bibitem{Likharev-1991} Likharev K K and Semenov V K 1991 \textit{IEEE Trans.Appl. Supercond.} {\textbf{1}} 3–28

\bibitem{Hazra-PRB-2010} Hazra D, Pascal L M A, Courtois H and Gupta A K 2010 \textit{Phys.Rev.B} \textbf{82} 184530

\bibitem{Tarte-2000} Tarte E J, Moseley R W, Kolbl M R, Booij W E, Burnell G and Blamire M G 2000 \textit{Supercond. Sci. Technol.} \textbf{13} 983

\bibitem{Skocpol-1974} Skocpol W J, Beasley M R and Tinkham M 1974 \textit{J. Appl. Phys.} {\textbf{45}} 4054–66

\bibitem{Courtois_2008} Courtois H, Meschke M, Peltonen J T and Pekola J P 2008
\textit{Phys. Rev. Lett.} {\bf 101} 067002

\bibitem{Biswas_2018} Biswas S, Winkelmann C B, Courtois H and
Gupta A K 2018 \textit{Phys. Rev. B}  {\bf 98} 174514

\bibitem{Barone} Barone A and Paterno G 1982 \textit{Physics and Applications of the Josephson Effect} (New York: Wiley)

\bibitem{CPR-review} Golubov A A, Kupriyanov M Yu and Il’ichev E 2004 \textit{Rev. Mod. Phys.} {\bf 76}, 411


\bibitem{Lam-2003} Lam S K H and Tilbrook D L 2003 \textit{Appl. Phys. Lett.} \textbf{82} 1078–80

\bibitem{Muck-1988} Mück M, Rogalla H and Heiden C 1988 \textit{Appl. Phys.A} \textbf{47} 285–9

\bibitem{Hadfield-2001} Hadfield R H, Burnell G, Booij W E, Lloyd S J, Moseley R Wand Blamire M G 2001 \textit{IEEE Trans. Appl.Supercond.} \textbf{11} 1126–9

\bibitem{Levichev-2019} Levichev M Yu, El’kina A I, Bukharov N N, Petrov Yu V, Aladyshkin A Yu, Vodolazov D Yu and Klushin A M 2019 \textit{Phys. Solid State} \textbf{61} 1544–1548

\bibitem{Vodolazov-2018} Vodolazov D Yu, Aladyshkin A Yu, Pestov E E, Vdovichev S N, Ustavshikov S S, Levichev M Yu, Putilov A V, Yunin P A, El'kina A I, Bukharov N N and Klushin A M 2018 \textit{Supercond. Sci. Technol.} \textbf{31} 115004

\bibitem{Kuprianov-JETP-1988} Kupriyanov M Yu and Lukichev V F 1988 \textit{Sov. Phys. JETP} \textbf{67} 1163

\bibitem{Baratoff-1970} Baratoff A, Blackburn J A and Schwartz B B 1970 \textit{Phys. Rev. Lett.} \textbf{25} 1096

\bibitem{Zubkov-1983} Zubkov A A and Kupriyanov M Yu 1983 \textit{Fiz. Nizk. Temp.} \textbf{9} 548

Zubkov A A and Kupriyanov M Yu 1983 \textit{Sov. J. Low Temp.Phys.} \textbf{9} 279  (Engl. transl.)

\bibitem{Vijay-2009} Vijay R, Sau J, Cohen M and Siddiqi I 2009 \textit{Phys. Rev. Lett.} \textbf{103} 087003

\bibitem{Baratoff-1975} Blackburn J A, Schwartz B B and Baratoff A 1975 \textit{J. Low Temp. Phys.} \textbf{20}, 523

\bibitem{Kupriyanov-LTP-1981} Kupriyanov M Yu and Lukichev V F 1981 \textit{Fiz. Nizk. Temp.} \textbf{7} 281

\bibitem{Perrin-1983} Perrin N and Vanneste C 1983 \textit{Phys. Rev.B} \textbf{28} 5150

\bibitem{Vodolazov_2017} Vodolazov D Yu 2017 \textit{Phys. Rev.
Appl.} \textbf{7} 034014

\end{thebibliography}
\end{document}